# GRID Architecture through a Public Cluster

Z. Akbar and L.T. Handoko

*Group for Theoretical and Computational Physics, Research center for Physics, Indonesian Institute of Sciences, Kompleks Puspiptek Serpong, Tangerang 15310, Indonesia*

Email : zaenal@teori.fisika.lipi.go.id, handoko@teori.fisika.lipi.go.id

**Abstract**
An architecture to enable some blocks consisting of several nodes in a public cluster connected to different grid collaborations is introduced. It is realized by inserting a web-service in addition to the standard Globus Toolkit. The new web-service performs two main tasks : authenticate the digital certificate contained in an incoming requests and forward it to the designated block. The appropriate block is mapped with the username of the block's owner contained in the digital certificate. It is argued that this algorithm opens an opportunity for any blocks in a public cluster to join various global grids.

## 1. Introduction

LIPI Public Cluster (LPC) is a unique, may be the first one in its kind [1,2], parallel machine which is open for public use [3]. The main difference is LPC provides full ownership to the user on a block of parallel machine consisting of several nodes using the so-called multi block approach [4]. On the contrary, the conventional "public" parallel machines allow the users to only put their jobs in the queues, and let the resource allocation management system like openPBS to distribute it appropriately according to the currently available resources [5]. In LPC, the users are granted with much higher degree of freedoms to control their own blocks, although all commands are executed through a web interface [6].

Microcontroller-based remote control and monitoring for hardwares in LPC makes it possible to be fully controlled by the users [7]. Since a user at a certain allocated period fully owns a block of nodes, the resource management system is irrelevant in LPC. Instead, the resource allocation is more likely needed to help the administrator to assign appropriate number and types of nodes for each incoming initial request according to the current availability and user needs [8]. This is the main reason in LPC we have removed the conventional resource allocation module [6,8].

On the other hand, advancing the LPC and connecting a block as a participating node in a global grid would require a tool to guide a request from partner grid to be forwarded properly to an appropriate block. In other word we have to define the queues available in LPC associated with the blocks. In this context we should still deploy the resource allocation management, but only to define the queues in LPC.

In this paper we discuss a compact architecture to connect any blocks in LPC to any global grids. The architecture makes use of available tools like some modules embedded in the Globus Toolkit (GT) [10], openPBS to set up the queues associated with each block [5]. Further, we develop a "router web service" to reroute the incoming request to an appropriate block by inserting a label on it containing the username of block's owner. Because in LPC each block is exclusively owned by a particular user.

## 2. Concept and architecture

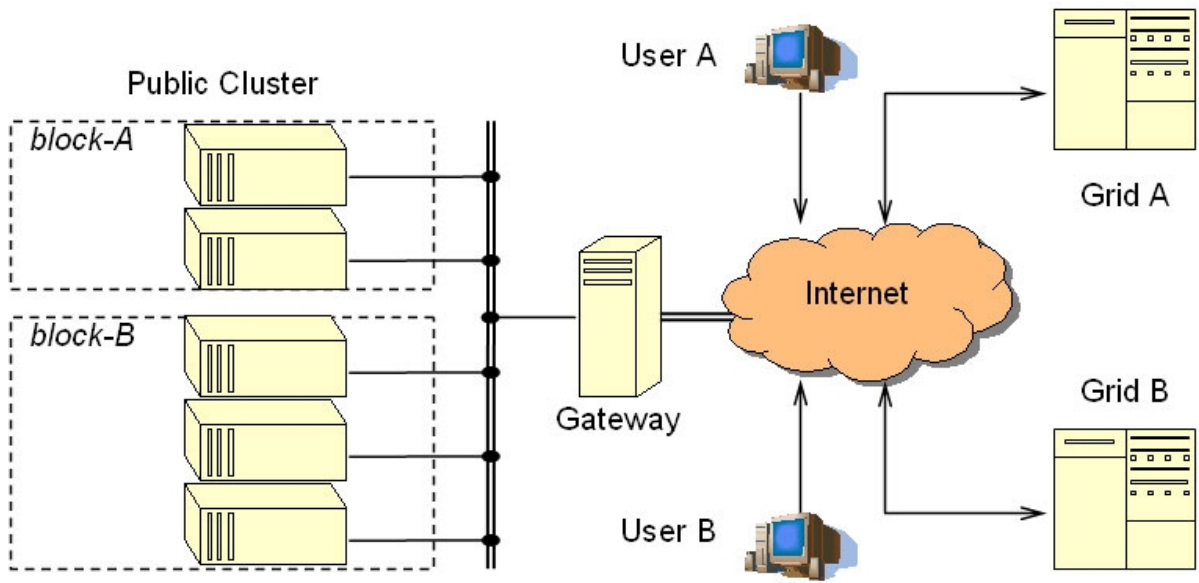

Fig. 1. A considerable case of connecting some blocks in LPC to different global grids.

Connecting a block in LPC to a grid simultaneously requires implementations of several things :
- A grid middleware. In our case, we borrow a widely used the Globus Toolkit version 4 (GT4) [10].
- A smart authentication method to reroute any requests from separate grids to appropriate blocks in LPC.

As mentioned above, due to its unique characteristic, in LPC there is originally no need to deploy a resource management tool at all. Because the whole nodes in a block is allocated for a single user who has full control for everything as if the user has their own parallel machine. Unfortunately, this rare concept leads to a severe problem once the block is going to be involved in a global grid. The problem is how to forward a request to the block appropriately. Moreover, there is also a considerable case when some blocks in LPC are connected to different grids as depicted in Fig. 1. Using a resource management tool in a conventional way would not resolve the problem, since all available nodes are already assigned in several independent blocks. Independent here means each of them is owned by different user, and may deploy various middlewares according to user needs.

How should we overcome this problem ? Also, how do we fulfill the needs while deploying the modules of GT as they are ? Exploiting the fact that a block in LPC is always occupied by a single user, we should deploy a resource management tool to define a queue name for the block. In our case we deploy openPBS for this purpose [5]. The problem is then how to specify each incoming request with a label to reroute it to the appropriate block through GT properly.

In order to overcome this problem we have developed a unique web service, namely Web Service Public Cluster (WSPC). WSPC plays an important role to authenticate the digital certificate contained in the requests, to retrieve the username and to map it with the queue name associated with particular block. The WSPC is placed beyond GT which also acts as certificate authority (CA) as usual. Since we deploy the GT as it is, double authentication occurs at WSPC and MyProxy of GT as well.

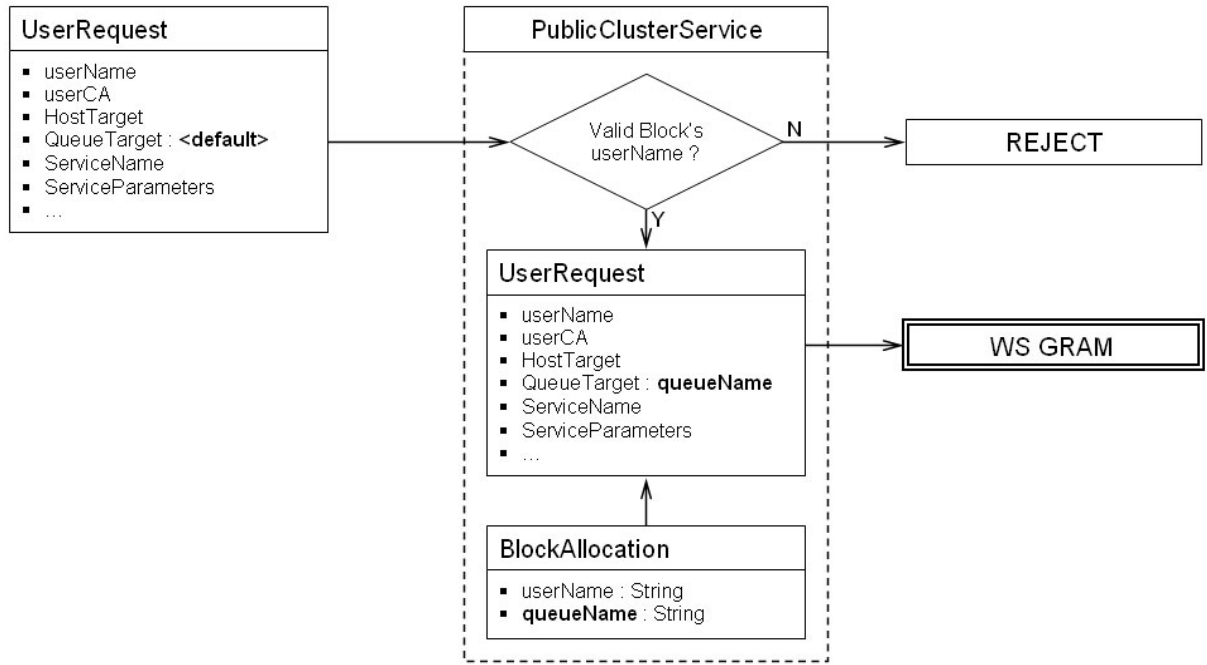

Fig. 2. The flow diagram for an incoming request from collaborating grid to a block through WSPC and WS GRAM of GT.

## 3. Implementation

Installing the whole GT4 and our WSPC, in principle, could provide smooth communication between the global grids and the participating blocks. More importantly, the blocks still have full flexibilities in the context of middleware for both parallel and grid environments. This is very crucial to keep LPC as an open infrastructure for public with varying needs and requirements.

The overall flow of these processes is shown in Fig. 2. In reality the implementation is done as follows :
1. The approved user is issued with a username and a set of nodes as a block. At the same time, a new queue is created with the same username for the allocated block. The queue is defined with specific nodes, and not allowed to use another ones. Also, the authorized user for the block is only the user which has just been approved..
2. When the user makes a request to his / her block, the following parameters should be sent together : username and userCA.
3. The queue name where the request will be forwarded to is determined by WSPC according to the containing username after mapping it with the data of available blocks at certain time.

Finally, both GT4 and WSPC are installed in the gateway of LPC. All components to handle the above-mentioned tasks are written in Fig. 3. We should remark here that we can deploy several grid middlewares which can be chosen by users. In order to distinguish all of them, each middleware is put in different directories. Hence the user request should follow the correct path of desired middleware according to its initial choice as activating the block.

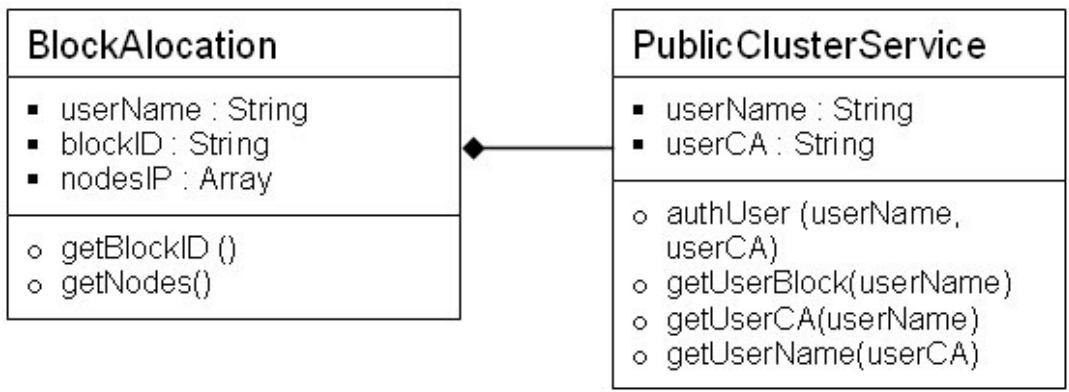

Fig. 3. The components of WSPC and its relation with the LPC.

**4. Summary**

We have introduced a new architecture for grid computing using public clusters with a characteristic like LPC. The architecture consists of the standard GT and an additional web service. The web service, WSPC, has a role as an intermediate interface to authenticate the digital certificate in a request, to retrieve the username and finally to map it with the queue name associated with particular block. It is argued that this architecture keeps the freedom and flexibility for users. Because different parallel and grid middlewares can be used simultaneously in separate blocks according to user needs.

As future work and issue, we are going to integrate the WSPC as a module or web service of GT4. This would improve the authentication process and avoid some delays due to double authentications. This can be achieved by removing MyProxy and embed some of its relevant features to the WSPC, or inversely adding unique functions of WSPC to MyProxy of GT4.

**Acknowledgment**

This work is financially supported by the Riset Kompetitif LIPI in fiscal year 2008 under Contract no. 11.04/SK/KPPI/II/2008.